\documentclass[letterpaper]{article} 
\usepackage{aaai23}  
\usepackage{times}  
\usepackage{helvet}  
\usepackage{courier}  
\usepackage[hyphens]{url}  
\usepackage{graphicx} 
\usepackage{natbib}  
\usepackage{caption} 
\usepackage{algorithm}
\usepackage{algorithmic}

\usepackage{newfloat}
\usepackage{listings}
\DeclareCaptionStyle{ruled}{labelfont=normalfont,labelsep=colon,strut=off} 
\floatstyle{ruled}
\newfloat{listing}{tb}{lst}{}
\floatname{listing}{Listing}
\usepackage{courier}

\title{Reducing the Environmental Impact of Wireless Communication via Probabilistic Machine Learning}

\author {
    A.~Ryo Koblitz,
    Lorenzo Maggi, 
    Matthew Andrews
}
\affiliations {
    Nokia Bell Labs\\
    \{\texttt{first.last}\}@nokia-bell-labs.com
}

\usepackage{bibentry}

\begin{document}

\maketitle

\begin{abstract}
    Machine learning methods are increasingly adopted in
    communications problems, particularly those arising in next
    generation wireless settings. Though seen as a key climate
    mitigation and societal adaptation enabler, communications related
    energy consumption is high and is expected to grow in future
    networks in spite of anticipated efficiency gains in 6G due to exponential
    communications traffic growth. To make meaningful climate mitigation impact
   in the communications sector, a mindset shift away from maximizing throughput at all
    cost and towards prioritizing energy efficiency is needed. Moreover, this
    must be adopted in both existing (without incurring further embodied carbon
    costs through equipment replacement) and future network infrastructure, given
    the long development time of mobile generations. To that end, we present
    summaries of two such problems, from both current and next
    generation network specifications, where probabilistic
    inference methods were used to great effect: using Bayesian
    parameter tuning we are able to safely reduce the energy
    consumption of existing hardware on a live communications network by 11\% whilst
    maintaining operator specified performance envelopes; through
    spatiotemporal Gaussian process surrogate modeling we reduce the
    overhead in a next generation hybrid beamforming system by over 60\%,
    greatly improving the networks' ability to target highly mobile users such
    as autonomous vehicles. The 
    Bayesian paradigm is itself helpful in terms of energy usage, since training a Bayesian
    optimization model can require much less computation than,
    say, training a deep neural network.
\end{abstract}

\section{Introduction}
\label{sec:introduction}
The sixth assessment report of the Intergovernmental Panel on Climate Change
(IPCC) makes for sobering reading: even in the best case scenario we are
likely to exceed +1.5~$^\circ$C global warming in the near future with 
far-reaching impacts spanning human and ecosystems \citep{RN14}. The 
information and communication technology (ICT) industry---in particular, 
the massive bandwidth and low latency
promised by next-generation network architecture---is seen as a key enabler 
for both societal adaptation and climate mitigation, 
e.g.~through expanded online services in healthcare,
education, and work, and through decentralized
renewables based microgrids \citep{I2020}.

However, as the popular adage goes, ``there ain't no such thing as free
lunch''. The ICT industry already accounts for
3\% of global energy consumption \citep{gsma} and an estimated
2-4\% of the world's greenhouse gas emissions (GHG) \citep{Freitag2021}. 
Networking makes up a
significant proportion of the ICT sector, with radio access networks alone (RAN)
accounting for over half of its energy consumption \citep{5Gsurvey}. All
the major RAN vendors (Ericsson, Huawei, Nokia, and Samsung) have committed to
ambitious net-zero targets---in many cases exceeding the voluntary
standard by
aiming for net-zero by as early as 2030 \citep{EricssonSustainability}---and are investing
heavily in developing next generation wireless as a climate mitigation
enabler \citep{HuaweiClimate,NokiaClimate,EricssonClimate}. Their customer base
(mobile network
operators), for whom energy costs make up a bulk of their operating expenditure~\citep{gsma},
are likewise prioritizing energy efficiency. Standards bodies are also bringing
energy efficiency to the forefront:
3GPP\footnote{Consortium of
major standards organizations which develop mobile telecommunications
protocols} 
have set a 90\% energy reduction goal for the next generation 
architecture New Radio (NR),
 and
machine learning will play an increasingly important role in unlocking next
generation wireless efficiencies~\cite{Elbir2023}. 
Regional and national future communication development initiatives 
have likewise placed sustainability at the forefront \cite{NextG, fonrc, hexax}.

Even 
though 5G and beyond will confer
substantial energy efficiency in terms of Joules per bit transferred,
in absolute terms consumption is predicted to increase~\citep{Freitag2021}. This is predicated on 
an anticipated
rebound effect---where increased efficiency and bandwidth give way to a surge in
demand---intrinsic 
higher network element density compared to LTE 
(of e.g.~mobile base stations, user devices), the need for network
operators to support mixed frequency sites for years to come, and embodied
carbon in the production of new RAN infrastructure. Moreover,
given
the long development time (typically 10 years between generations) and
disparate rollouts (5g will not be the dominant service until after 2028, and
even then, many regions---notably sub-Saharan Africa---will still rely heavily on
previous generation hardware \citep{EricssonMobility}), it is
imperative that both existing and future infrastructure is optimized for
efficiency. This will require a mindset shift away from maximum
performance to sustainable performance, where throughput and energy
consumption are jointly optimized. Energy efficiency may
be addressed at various layers in the RAN architecture,
from physical layer processing, to network-scale radio resource 
control processing. Here we focus on higher level
radio resource management (RRM) applications, for examples of ML applied to the physical layer see
e.g.~\citet{Korpi2021} and references therein. In this paper we will 
present summaries of two
applications of probabilistic machine learning to increase energy efficiency in
RAN. 
Crucially, we consider both current generation 3GPP LTE and next
generation 3GPP New Radio (NR). 
Both settings utilize Bayesian inference
which is itself helpful in terms of energy usage as it typically requires far
less computation than, say, training a deep neural network.

\section{Learning considerations}
\label{sec:preliminaries}
RRM problems are challenging to optimize, with objectives that tend 
to be expensive to evaluate either through simulations or via point estimates 
of user quality of service \citep{Zhang2023}. Many approaches leverage deep 
neural network based techniques, with
the goal of developing generalized models for wide deployment. These approaches
can have good performance at inference time with relatively low energy
requirements. However, they are typically expensive (from a computational and
therefore energy point of view) to train. Another practical consideration is
model life-cycle management: periodic re-training/tuning and model transfer
over-the-air from base station to user equipment are challenges that should not
be underestimated.

Both applications considered in this paper share
characteristics--namely, low
dimensionality and information scarcity--making them amenable to 
online Bayesian optimization. Bayesian optimization is a
class of stochastic optimization methods suitable for derivative-free
optimization of expensive to evaluate black-box objective functions, see
\citet{Frazier} for an excellent introduction. There are
two core aspects: a surrogate model, developed using Bayesian statistics, 
of the true objective, and an acquisition policy, which uses the surrogate
model to infer where to sample the objective next. Because we forego 
generalization across wide ranging network deployments,
and our problem dimensions are small, the costly Gaussian process parameter
fitting\footnote{For us this involves a matrix inversion with memory
requirements that scales cubically in the number of data points.} is abated,
making this a low-cost alternative to wide deployment generalized neural
network based techniques.

\section{Results and Discussion}
We now describe the cellular energy efficiency problems in more
detail. For current networks we are typically faced with a situation
where traffic at a cell site fluctuates during the day. In this case it
is wasteful to keep all carriers (i.e.\ frequencies) and their
associated power amplifiers active at all time. We therefore want to
learn a strategy that reacts to the changing traffic and switches off
carriers to save energy whenever this can be done without violating
customer performance metrics.

For future networks, large antenna arrays improve efficiency by
focusing energy at the intended receiver. For this mechanism to work,
the cell must track the best beam for each mobile user, and it is
prohibitively expensive to measure every beam at every time
instant. However, if the wrong beam is selected then energy is wasted
because the focused beam ``misses'' its target. We therefore apply
Bayesian techniques to learn the best beam with a minimal number of
measurements.

\subsection{Current generation: cell switch-off}
\label{subsec:es}
\begin{figure}[h]
    \centering
    \includegraphics[width=0.48\linewidth]{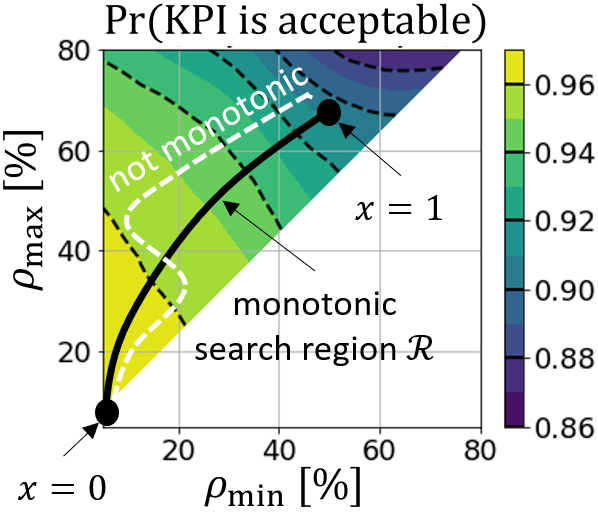}
    \includegraphics[width=0.48\linewidth]{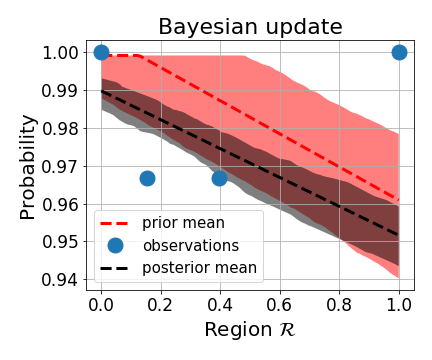}
    \caption{Left: Adjusting load thresholds ($\rho$) to effect carrier switch-off 
    is transformed to a Bayesian root finding problem. The search space (density plot 
    of example quality of service metric) is reduced to a monotonic region $\mathcal{R}$.
    Right: The parameterized probability function of our root finding problem 
    is refined using Bayesian posterior updating. Figure adapted from~\citet{Maggi1} \copyright~2023 IEEE.}
\end{figure}

Power amplifiers (PA) account for over 65\% of a base stations energy
consumption \citep{5Gsurvey}. Switching off PAs without (overly) degrading
network performance is therefore a good route to curbing energy consumption.

In \citet{Maggi1} 
we develop an approach whereby carriers are shut down according to a
hysteresis mechanism: the next carrier in line is switched off (on) if the 
traffic load on the sector is lower (greater) than a certain
threshold $\rho_\mathrm{min}$ ($\rho_\mathrm{max}$). The key is dynamic selection
of the thresholds such that key performance indicators (KPIs) meet quality of
service (QoS) constraints, with a desired likelihood. We formulate this as a 
root finding problem, in which the solution is mapped to a one-dimensional
space that is iteratively searched using a Bayesian approach, see
\citet{Maggi1} for derivation and implementation details. Briefly, we
parameterize a probability function of our threshold, describing the
probability of satisfying service constraints, and maintain a probabilistic
belief over the parameter space. We convolve this belief with a Markovian transition
law, allowing traffic changes to be learned, and update this belief with new
observations via Bayesian posterior updating.

We tested this policy in a proof of concept (PoC) trial 
on a live customer 4G network, 
spanning 19 sites and 57 sectors. Most
of the sites had 4 frequency layers (800, 1800, 2100 and 2600 MHz), with the
800 and 1800 MHz layers left active to preserve coverage. Baseline measurements
were taken over periods spanning a few weeks immediately before and after the
PoC trial, during which all carriers were kept active. We used two weeks
of historical data to initialize the prior, and we used
a Gaussian distribution with zero mean and diagonal covariance matrix as our
Markovian transition rule, allowing the
policy to adapt to traffic variations by gradually ``forgetting'' past
observations. During our PoC, we were able to shut down carriers for
approximately 30\% of the time with negligible impacts on cell congestion and  
traffic volume, including cells in neighboring sites.
Overall, we are able to effect an 11\% reduction in energy consumption during
the PoC, which is remarkable considering no hardware changes were made.

\subsection{Next generation: beam tracking}
\label{subsec:bt}
\begin{figure}
    \centering
    \includegraphics[width=0.8\linewidth]{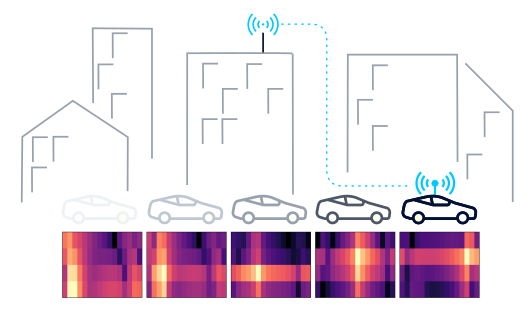}
    \includegraphics[width=\linewidth]{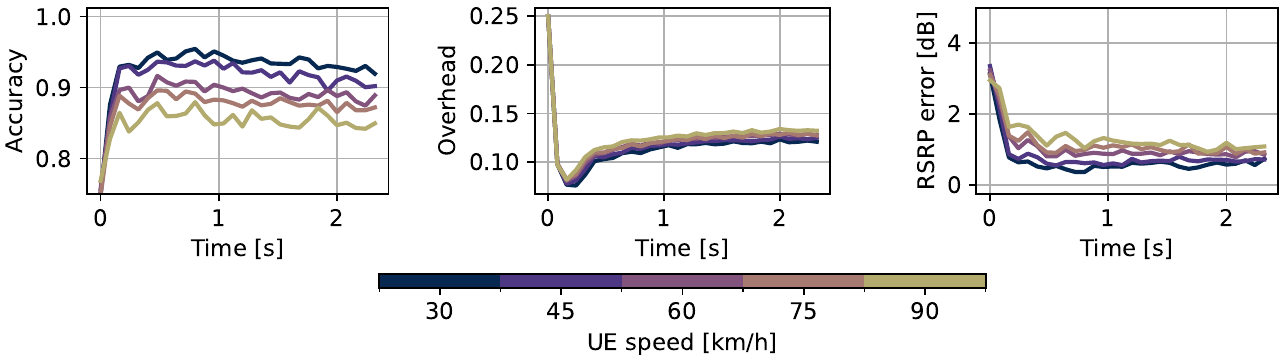}
    \caption{Highly mobile UEs, such as connected vehicles, will
    experience 
    different radio environments over time, requiring tracking of the best
    transmission and receiver beam pairs. The schematic (top) shows
    a mobile UE traversing a cell sector in an urban environment, maintaining a
    communication link with a base station. As the UE moves from left to right,
    the radio environment changes (due to, for example, radio reflections),
    which affects the signal strength of the base station beams, as experienced
    by the UE. This is indicated by the contour plots below the UE (center),
    with brighter color indicating a stronger received signal reference power (RSRP),
    and therefore better beam choice for the communication link. The time
    series (bottom) show the efficacy of our Bayesian optimization solution in
    tracking the best beam for mobile UEs, with varying mobility level
    indicated by the color (brighter is faster): the online learning approach
    suffers a cold start of less than 300\,ms, with minimal degradation of RSRP
    error and overhead metrics for increasingly mobile UEs. Figure adapted from~\citet{Maggi2} \copyright~2023 IEEE.}
\end{figure}
Next generation mmWave communications will use highly directional transmitter
and receiver (Tx and Rx, respectively) beams, complicating the mutual discovery 
process--both in initial access (IA) and the recovery from link failure--
and the ongoing beam tracking for highly mobile UEs such as connected vehicles
or mobile users on mass transit. Finding a good beam pair is
important not only from a QoS point of view but also energy efficiency:
a suboptimal beam pair means that the Tx and Rx beams are misaligned, so energy
is wasted by effectively pointing beams at empty space. Even a loss as
small as 1 dB equates to approximately 20 percent power waste in transmission,
so as our wireless communications become more directional and highly mobile
UEs more ubiquitous efficient beam alignment will go a long way
towards curbing emissions growth.

\begin{figure}
    \centering
    \includegraphics[width=\linewidth]{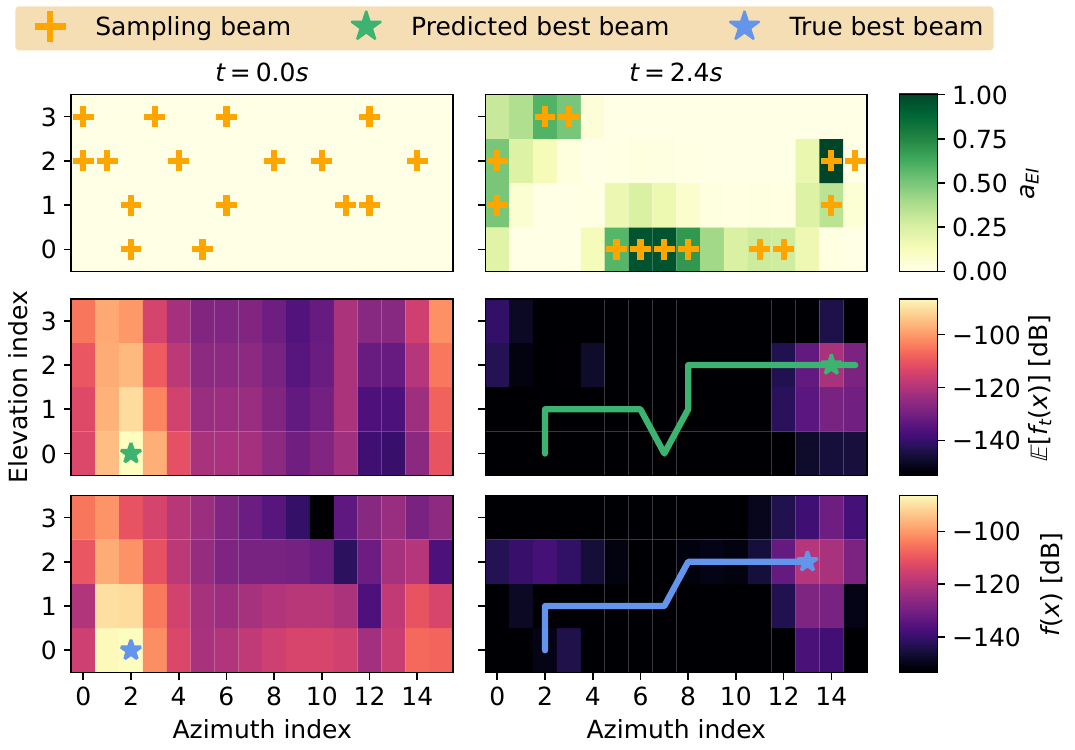}
    \caption{We visualize the salient components of our beam tracking solution 
    at an initial and late stage time slot. Top row: with an uninformative
    prior, we distribute initial beam samples (orange crosses) according to a
    space filling sequence (left). At subsequent time slots (right) samples are
    distributed according to an acquisition function (darker is better). Bottom
    row: The ground truth signal strength (brighter is better) experienced by
    the UE for each transmission beam index changes as the time slots evolve.
    The optimal beam index is denoted with a blue star, and the solid line traces
    the evolution of the optimal beam index. Center row: we plot the
    expectation over our posterior estimate of the signal strength across beam
    indices. This estimate is used to predict the optimal beam index for a
    given time slot (green start), and once again the solid line traces the
    evolution of this predicted optimal beam index. Figure adapted from~\citet{Maggi2} \copyright~2023 IEEE.}
\end{figure}

The signal received by a UE is a function of its gain, exogenous noise, the
selected Tx and Rx beams, and the channel state information
(CSI\footnote{Complex valued matrix whose rank depends on BS and UE antenna
array sizes}) matrix. 
A good signal has high magnitude, known as the received signal 
reference power (RSRP). If the CSI is known, 
then the RSRP can be computed for all Tx-Rx beam pairs, making beam 
selection trivial. However, if no CSI is available, then beam pair 
selection must be made based on RSRP measurements. Current standards employ an
exhaustive search whereby all beam pairs are sampled and the best one selected.
Because beam width is inversely proportional to array size, and analog 
beam formers may only probe one direction at a time, this exhaustive search is
significantly more expensive for next generation architectures that use far
larger antenna arrays.

In \citet{Maggi2} we developed an alignment and tracking procedure using
Bayesian optimization to greatly reduce the overhead required during both IA
and subsequent beam tracking, especially when dealing with highly mobile UEs.
We maintain a Gaussian process surrogate model of the RSRP landscape over all
beam pairs, and use this model with a modified expected improvement based
acquisition function to rapidly identify the best beam pair in a given time
slot. Through appropriate Gaussian Process (GP) kernel design we can cater for highly mobile
UEs, modeling the RSRP evolution in both space and time. See \citet{Maggi2} for
modeling and implementation details, including a strategy for restricting the
size of the set of sampled beams with proven optimality gap.

We evaluated our approaching using a 3GPP NR-compliant system level simulator,
and benchmarked its performance against exhaustive search. A full system
description and complete benchmark
comparisons against spatial interpolation and long short term memory based
approaches may be found in \citep{Maggi2}. Overall, we were able to use 60\%
fewer measurements than the current standard, resulting in more efficient use
of transmission power, without compromising alignment accuracy: RSRP errors
were kept around 1dB (translating to
approximately milliwatt power wastage due to beam misalignment), even for UEs
travelling at 90 km/h (highway speed). Importantly, the performance was found
to be stable over time, with no overhead surges or accuracy degradation evident.

\subsection{Conclusion}
Machine learning will play an important role in unlocking efficiencies in
current and next generation wireless networks---a key climate mitigation and
societal adaptation enabler. But to mitigate against growing ICT emissions an
industry mindset shift is needed to jointly optimize performance and
efficiency. Crucially, given the long development time of next generation
wireless networks, we must include current generation architectures in this
joint optimization effort to have an impact on ICT climate mitigation today.

In
this paper we introduced two problems, one from the current network generation,
the other from the next generation, where we applied probabilistic machine
learning to strike a balance between performance and energy consumption. We
sought to introduce radio resource management to the wider machine learning
community, whose contribution to the adoption of machine learning to next
generation wireless will be crucial.

\section{Acknowledgments}
This work is a contribution by Project REASON, a UK Government funded project 
under the Future Open Networks Research Challenge (FONRC) sponsored by the 
Department of Science Innovation and Technology (DSIT). 
We would like to thank the anonymous reviewers for their constructive feedback, 
and our colleague Mohammad Malekzadeh for commenting on an earlier version of this paper.

\bibliography{koblitz_aaai23}

\end{document}